\documentclass[]{GeotechAuth}

\providecommand{\tightlist}{%
	\setlength{\itemsep}{0pt}\setlength{\parskip}{0pt}}
\usepackage{newtxtext,newtxmath}
\usepackage[colorlinks,bookmarks=false,citecolor=red,urlcolor=red]{hyperref}
\usepackage{subcaption}

\usepackage{graphicx}
\makeatletter
\newsavebox\pandoc@box
\newcommand*\pandocbounded[1]{
	\sbox\pandoc@box{#1}%
	\Gscale@div\@tempa{\textheight}{\dimexpr\ht\pandoc@box+\dp\pandoc@box\relax}%
	\Gscale@div\@tempb{\linewidth}{\wd\pandoc@box}%
	\ifdim\@tempb\p@<\@tempa\p@\let\@tempa\@tempb\fi
	\ifdim\@tempa\p@<\p@\scalebox{\@tempa}{\usebox\pandoc@box}%
	\else\usebox{\pandoc@box}%
	\fi%
}

\begin{document}

\runningheads{Rotational Penetration}{Y. Tang, Y. Zhong and J. Tao}

\title{Experimental Study on the Rotation-induced Reduction of
	Penetration Resistance in Sand}

\author{
	Yong Tang\addressnum{1},
	Yi Zhong\addressnum{2} \authorand
	Junliang Tao\addressnum{3}
}


\address{
	\addressnum{1}
	Sustainable Engineering and the Built Environment, Center for
	Bio-medaited and Bio-inspired Geotechncis, Arizona State
	University, Tempe, AZ, USA. (Currently with Northwestern University)
	(Email: \href{mailto:ytang116@asu.edu}{ytang116@asu.edu})\\
	\addressnum{2}
	Sustainable Engineering and the Built Environment, Center for
	Bio-medaited and Bio-inspired Geotechncis, Arizona State
	University, Tempe, AZ, USA. (Currently with Geosyntec Consultants)
	(Email: \href{mailto:yzhong53@asu.edu}{yzhong53@asu.edu})\\
	\addressnum{3}
	Sustainable Engineering and the Built Environment, Center for
	Bio-medaited and Bio-inspired Geotechncis, Arizona State
	University, Tempe, AZ, USA
	(Corresponding author; Email: \href{mailto:jtao25@asu.edu}{jtao25@asu.edu})(ORCID: \href{https://orcid.org/0000-0002-3772-3099}{0000-0002-3772-3099})\\
}
\begin{abstract}
	Soil-dwelling organisms have evolved diverse strategies for efficient
	subterranean movement. For example, the seeds of \emph{Erodium
		cicutarium} and \emph{Pelargonium} species employ continuous rotational
	motion for self-burial, while the angled worm lizard (\emph{Agamodon
		angeliceps}) tunnels by oscillating its head around its trunk's axis.
	These rotational movements significantly reduce penetration resistance.
	This study presents comprehensive experiments investigating the effects
	of various factors on rotational penetration forces and energy
	consumption. Results reveal that force reduction follow an approximately
	hyperbolic decay with the tangential-to-axial velocity ratio (\(u\)).
	Penetrator geometry, particularly roundness and conical tip shape, is
	found to significantly influence reduction at low velocity ratios,
	whereas relative density and material type exhibit moderate impact.
	Reduction is also observed to increase with interfacial friction angle
	but decreases with confining pressure and depth. Energy consumption
	analysis shows that while penetration force-related energy decreases
	with \(u\), total energy consumption increases due to rotational torque.
	For self-burrowing robot designs, lower velocity ratios are recommended
	to balance penetration force reduction and energy efficiency
	effectively.
\end{abstract}


\maketitle
\section{Introduction}\label{introduction}

As one of the particular topics of bio-inspired geotechnics,
bio-inspired self-burrowing mechanisms and robots have attracted
significant attention in recent research
\citep{THT20, CKMD21, TT22b, MDA+22, BSB+23}. These self-burrowing
robots hold potential for applications in subsurface exploration,
monitoring, surveillance \citep{Tao21}, extraterrestrial sampling
\citep{WZZ+21}, and precision agriculture \citep{LMD+23}. However,
designing such robots poses considerable challenges due to the
complexity of their operating environments \citep{DD23}. Navigating
through soil is particularly difficult due to anisotropic stress states,
inhomogeneity, and multiphasic properties \citep{Tao21}. Additionally,
soil imposes significantly higher resistance than those encountered in
air or water \citep{Tao21, NKM+21}.

Despite this high resistance encountered in soil, various organisms have
evolved to burrow effectively and efficiently using diverse locomotion
strategies. These strategies include body undulation, as observed in
sandfish lizards (\emph{Scincus scincus}), worms (\emph{Armandia
	brevis}), and the burrowing eel (\emph{Pisodonophis boro})
\citep{MDLG09c, MDU+11, HCD+11, DLR13}; dual-anchor and fluidization
techniques utilized by razor clams \citep{TBD66a, Tru67, Tru66a, HT20a};
peristaltic crawling typical of earthworms \citep{Qui00, CUZP16a}; and
circumnutation movements seen in plant roots \citep{MTF13b, TLM+21a}.
The overarching principle behind these strategies is to either minimize
penetration resistance or to enhance anchorage or propulsion force
during the burrowing process, showcasing nature's varied approaches to
overcoming the challenges of soil navigation.

\begin{figure*}

	\centering{

		\pandocbounded{\includegraphics[keepaspectratio]{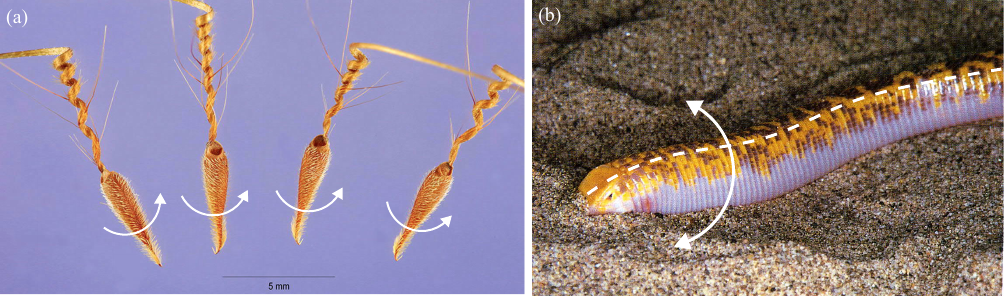}}

	}

	\caption{\label{fig-biomodel}Biological models: (a) seed awns of
		\emph{Erodium crinitum} \citep[modified from][]{PCM12}; (b) angled worm
		lizards, \emph{Agamodon angeliceps} \citep[modified from][]{CZK98}. The
		dashed arrows in panel (a) indicate continuous rotational movements,
		while the solid arrow in panel (b) represents oscillatory motion. The
		dashed line in panel (b) denotes the long axis of the trunk.}

\end{figure*}%

Seeds of \emph{Erodium} and \emph{Pelargonium},
Figure~\ref{fig-biomodel} (a), can bury themselves into the ground for
future germination through rotation induced by hygroscopic coiling and
uncoiling of their awns \citep{EGF08, ATK+12, AE13}. Similarly, angle
worm lizards, \emph{Agamodon anguliceps}, Figure~\ref{fig-biomodel} (b),
create tunnels by oscillating their heads around the long axis of their
trunks \citep{Gan68, Gan74}. Both organisms use rotational motions
(continuous rotation, oscillation) to reduce penetration resistance
during their penetration and burrowing process.

Recent studies have confirmed that rotational motion can reduce
penetration resistance through various approaches, including numerical
simulations, physical experiments, and theoretical formulations. For
instance, Deeks \citeyearpar{Dee08} and co-authors \citeyearpar{DW08}
experimentally demonstrated that the base resistance of a rotary-jacked
pile decreases under centrifuge conditions. In addition to these
centrifuge-based studies, rotational penetration experiments have been
conducted under 1g conditions
\citep{JKK14b, JCKK17c, GDB18, HI21, SMK20, SFA21}. Furthermore,
rotational penetration tests have been both simulated numerically under
elevated gravitational conditions \citep{SBC+21} and 1g conditions
\citep{PG12, TT22b, YZW+24}, examining different tangential-to-axial
velocity ratios. Specifically, with Discrete Element Method, Tang and
Tao \citeyearpar{TT22b} demonstrated that rotation-induced reduction of
penetration resistance are caused by reduced number of particles in
contact with the penetrator, lower magnitude of contact forces, and
reorientation of the contact forces away from vertical directions.
Additionally, the numerical results challenged conventional assumptions
from two theoretical models, indicating that normal stress on the tip
decreases with rotational velocity \citep{BMW97, SBC+21}.

Despite the many studies on rotation-induced reduction of penetration
resistance, the reported reduction percentages vary significantly across
investigations. This study aims to clarify and address these variations
through holistic experiments under various conditions. The experimental
methods are outlined, including material properties, penetrator setups,
procedures, and data analysis techniques. Results are presented for
penetration force, torque, force reduction, and energy consumption under
different conditions. The discussion explores the reasons behind the
varying reduction percentages, implications for energy consumption, and
study limitations. Finally, key findings and insights are highlighted.

\section{Methodology}\label{methodology}

\subsection{Materials}\label{materials}

\begin{figure}

	\centering{

		\pandocbounded{\includegraphics[keepaspectratio]{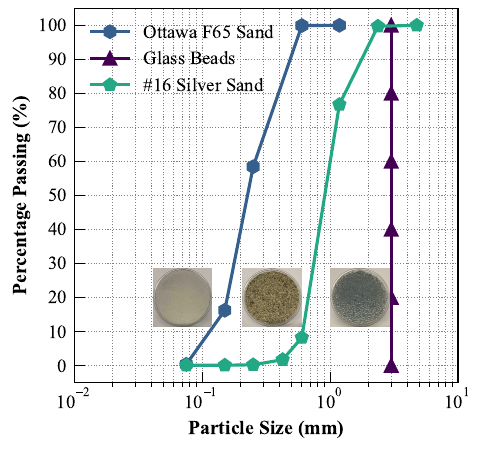}}

	}

	\caption{\label{fig-psd}Particle size distributions for different
		media.}

\end{figure}

This study used three types of granular media: Ottawa F65 sand, \#16
silver sand (a local sand), and glass beads. The particle size
distributions for these media are illustrated in Figure~\ref{fig-psd},
with inset figures highlighting the differences among them.
Table~\ref{tbl-materials} summarizes the key physical parameters of
these materials, including the mean particle size \(D_{50}\), the
coefficients of curvature (\(C_c\)) and uniformity (\(C_u\)), and the
maximum (\(e_\text{max}\)) and minimum (\(e_\text{min}\)) void ratios.
All media were used in shallow rotational penetration tests, while only
Ottawa F65 sand was used for tests involving deeper penetration depths
and confining pressures.

\begin{table*}

	\caption{\label{tbl-materials}Physical parameters for different media.}

	\centering{

		\centering

		\begin{tabular}{l|l|l|l|l|l|l|l}
			\hline
			Materials        & $D_{50}$ (mm) & $C_{c}$ & $C_{u}$ & $e_{\text{min}}$ & $e_{\text{max}}$ & Shape       & Classification \\
			\hline
			Ottawa F65 sand  & 0.23          & 1.06    & 2.26    & 0.79             & 0.56             & sub-rounded & SP             \\
			\#16 silver sand & 0.91          & 0.91    & 1.64    & 0.85             & 0.57             & sub-angular & SP             \\
			glass beads      & 3             & 1       & 1       & 0.64             & 0.55             & spherical   & -              \\
			\hline
		\end{tabular}

	}

\end{table*}%

\subsection{Penetrators and Experimental
	Setups}\label{penetrators-and-experimental-setups}

\begin{figure*}

	\centering{

		\pandocbounded{\includegraphics[keepaspectratio]{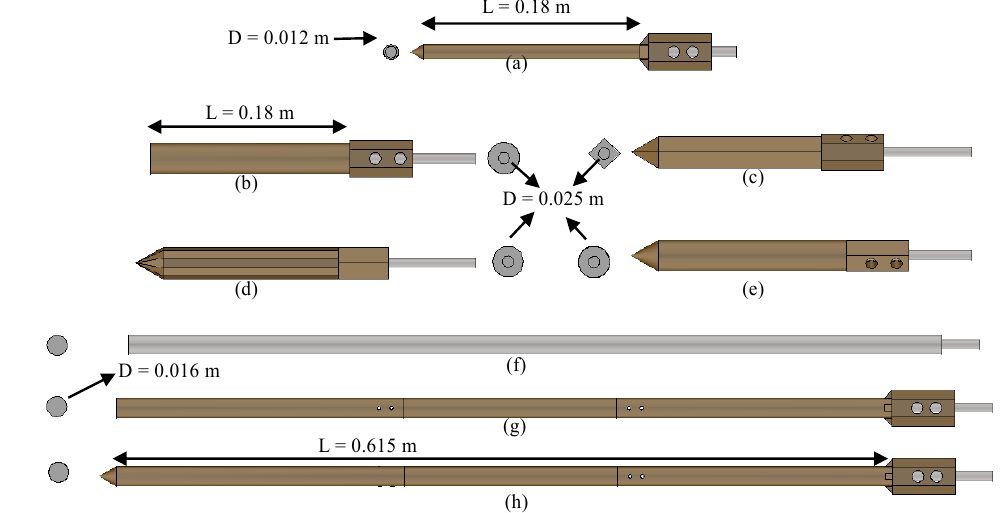}}

	}

	\caption{\label{fig-penetrator}Penetrator geometries: (a) confining
		pressure condition; (b) to (e) shallow condition; (f) to (h) deep
		condition.}

\end{figure*}%

Penetrators in this study were categorized into three groups based on
their diameters, as shown in Figure~\ref{fig-penetrator}. The diameters
are 25 mm, 16 mm, and 12 mm for penetration tests under shallow, deep,
and confining pressures conditions, respectively. The choice of
penetrator diameters was influenced by the load capacity of the robotic
arm (Universal Robotics UR16e with a payload of 16 kg) and the use of a
standard 16 mm stainless steel rod, Figure~\ref{fig-penetrator} (f).
Except for the steel rod penetrator, all other penetrators were 3D
printed using Polylactic Acid (PLA) material. The penetrators typically
have circular cross-sections, with the exceptions of
Figure~\ref{fig-penetrator} (c) and (d), which feature square and
decagon shapes, respectively. However, the cross-sectional areas of
Figure~\ref{fig-penetrator} (b) to Figure~\ref{fig-penetrator} (e) have
the same circumscribed circle with a diameter of 25 mm. Additionally,
the apex angle for all penetrators is 60°. Each 3D printed penetrator
was attached to a rotary motor (Pololu 4847) through customized
couplers.

\begin{figure}

	\centering{

		\pandocbounded{\includegraphics[keepaspectratio]{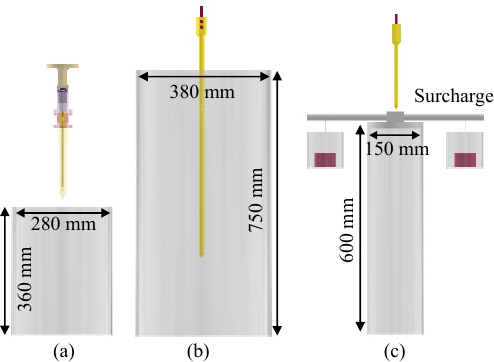}}

	}

	\caption{\label{fig-container}Experimental setups: (a) shallow
		condition; (b) deep condition; (c) confining pressure condition.}

\end{figure}

The schematic diagrams in Figure~\ref{fig-container} illustrate the
experimental setups and dimensions for rotational penetration tests
under various conditions. Notably, the setup for tests with confining
pressures, Figure~\ref{fig-container} (c), features a rigid plate on top
of the soil chamber, distinguishing it from the shallow and deep test
setups. Specifically, two steel rods are integrated into the rigid
plate, allowing for the addition of surcharges via small containers on
either side.

\subsection{Experimental Procedures and
	Plans}\label{experimental-procedures-and-plans}

All granular samples were prepared using the dry air pluviation method.
For dense samples under shallow conditions, vibration with a shaker
table was applied after pluviation. Free dry air pluviation consistently
achieved a relative density of around 35\%, while for dense samples, the
relative density was controlled to be 85\%. For samples under confining
pressure, in addition to the air pluviation and densification processes,
surcharges were placed on both sides of the container and maintained for
thirty minutes to ensure complete load transfer to the bottom, as
illustrated in Figure~\ref{fig-container} (c). A hole slightly larger
than the penetrator was made in the 3D-printed plate to accommodate the
penetration process. The confining pressure tests were conducted with
the surcharges in place throughout.

Several steps were undertaken to control the robotic arm and rotary
motor during each rotational penetration test. Initially, both the
robotic arm and rotary motor were activated to facilitate the vertical
and rotational movements separately. Next, the force/torque sensors of
the robotic arm were zeroed before the penetrator approached the sample
surface. The robotic arm was then paused for 3 seconds to start the data
acquisition system. The penetrator proceeded to penetrate the sample at
preset constant rotational and vertical velocities, stopping either at a
predefined penetration depth or upon reaching the maximum force
threshold. Lastly, the penetrator was retracted to its original position
through reverse motion. Further details on the experimental procedures
are available in \citep{TZTa}.

\begin{figure*}

	\centering{

		\pandocbounded{\includegraphics[keepaspectratio]{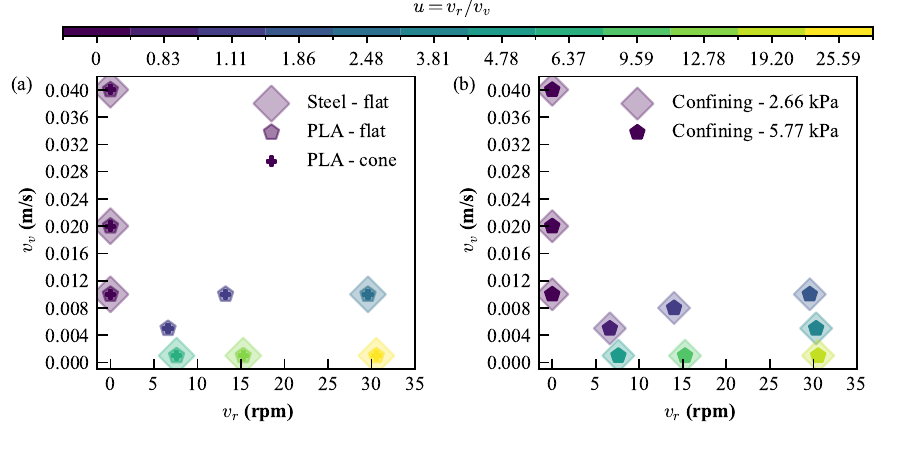}}

	}

	\caption{\label{fig-plansdeepconfining}Experimental plans for different
		conditions: (a) deep conditions; (b) confining pressure conditions. (The
		points at the same locations corresponding to the same \(u\) are shown
		with markers. To differentiate between various cases, the brightness of
		the markers is adjusted slightly. The PLA-cone and 5.77 kPa cases retain
		the original brightness of the colorbar.)}

\end{figure*}

Table~\ref{tbl-shallow}, Figure~\ref{fig-plansdeepconfining} (a) and
Figure~\ref{fig-plansdeepconfining} (b) summarize the experimental plans
for rotational penetration tests under shallow, deep, and confining
pressure conditions, respectively. For the shallow condition,
twenty-nine groups of experiments were conducted, varying granular
media, relative densities, and penetrator geometries. Each group
consisted of tests with different combinations of vertical and
rotational velocities, maintaining the same resultant velocity,
following the methodology outlined in \citep{TZTa}. For the deep
condition, twenty-five groups of tests were conducted using various
penetrators, each differing in geometry and interface features. For the
confining pressure condition, twenty groups were tested with the same
penetrator. Both the deep and confining pressure tests were conducted in
a loose state. For all conditions, each test was conducted three times
to ensure repeatability. In total, 585 tests were conducted in this
study.

\begin{table}

	\caption{\label{tbl-shallow}Experimental plans for shallow rotational penetration tests}

	\centering{

		\centering

		\begin{tabular}{p{1.8cm}|p{1.2cm}|p{1.5cm}|p{2.2cm}}
			\hline
			Materials   & Densities & Penetrators & $v_{v}$(m/s)                          \\
			\hline
			Ottawa F65  & Loose     & Decagon     & 0.01, 0.02, 0.04, \newline 0.08, 0.10 \\
			Glass beads & Loose     & Decagon     & 0.04, 0.08, 0.10                      \\
			\#16 silver & Loose     & Decagon     & 0.04, 0.08, 0.10                      \\
			Glass beads & Loose     & Flat-end    & 0.04, 0.08, 0.10                      \\
			Glass beads & Loose     & Square      & 0.04, 0.08, 0.10                      \\
			Glass beads & Loose     & Circular    & 0.04, 0.08, 0.10                      \\
			Ottawa F65  & Loose     & Flat-end    & 0.04, 0.08, 0.10                      \\
			Glass beads & Dense     & Circular    & 0.04, 0.08, 0.10                      \\
			Ottawa F65  & Dense     & Decagon     & 0.01, 0.02, 0.04                      \\
			\hline
		\end{tabular}

	}

\end{table}%

\subsection{Data Analysis}\label{data-analysis}

\subsubsection{Penetration force ratio}\label{penetration-force-ratio}

\begin{figure*}

	\centering{

		\pandocbounded{\includegraphics[keepaspectratio]{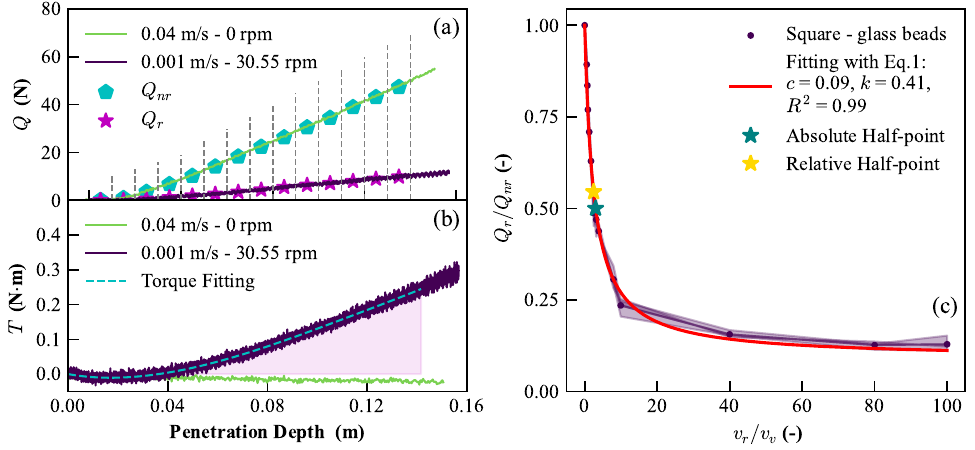}}

	}

	\caption{\label{fig-analysisexample}Analysis methods illustration: (a)
		penetration force reduction; (b) torque energy; (c) hyperbolic curve
		fitting and illustration of half points.}

\end{figure*}

Figure~\ref{fig-analysisexample} (a) illustrates the method used to
calculate the penetration force ratio \(Q_{r}/Q_{nr}\), or relative
penetration force, through a three-step process:

\begin{itemize}
	\tightlist
	\item
	      For each 1 cm interval in penetration depth, the mean force was
	      calculated for all trials in a rotational penetration case (\(Q_{r}\))
	      and the control direct penetration case (\(Q_{nr}\)).
	\item
	      The \(Q_{r}\) was then normalized by the corresponding \(Q_{nr}\) to
	      determine the relative penetration force at each depth interval.
	\item
	      Finally, the overall penetration force ratio was calculated by
	      averaging the ratios across all depth intervals.
\end{itemize}

Theoretically, using force ratios between rotational and direct
penetration cases to represent penetration force reduction is imprecise
due to differing vertical velocities for the rotational and control
cases in the experiments. However, experimental findings by Kang
\emph{et al.} \citeyearpar{KFLB18a} demonstrated that the axial drag
force remains nearly constant in a quasi-static soil state. Similarly,
\citet{Rot21} and \citet{RHJ21} observed that the velocity dependence of
drag force occurs only immediately after impact, with drag forces
stabilizing across all intrusion speeds, from the quasi-static to the
dense flow regime, beyond a crossover point. Thus, reductions were
directly calculated by comparing the ratios between the rotational cases
and their corresponding control cases, based on these assumptions.

\subsubsection{Reduction curve}\label{reduction-curve}

The tangential-to-axial velocity ratio, or velocity ratio for
simplicity, is defined as the ratio between the tangential (rotational)
velocity \(v_r\) and the axial (vertical) velocity \(v_v\), or
\(u = v_r/v_v = \omega D/2v_v\), where \(\omega\) are the angular
velocity and \(D\) is the diameter of the penetrator.

The force ratio \(Q_{r}/Q_{nr}\) can be plotted against the velocity
ratio to reveal the trends in the reduction of penetration force
(Figure~\ref{fig-analysisexample} (c)). This reduction curve can be
fitted using a hyperbolic decay equation, as shown in
Equation~\ref{eq-fit}, and represented by the red line in
Figure~\ref{fig-analysisexample} (c).

\begin{equation}\phantomsection\label{eq-fit}{ Q_{r}/Q_{nr} = c + \frac{1-c}{1+ku} }\end{equation}

where \(c\) and \(k\) are the fitting parameters, with \(k\) indicating
how quickly \(Q_{r}/Q_{nr}\) ratio decays with velocity ratio and \(c\)
representing the ultimate reduction achievable at infinite \(u\). Also
note that the initial ratio of \(Q_{r}/Q_{nr}\) is always unity when
\(u=0\), or without rotation.

From the fitted curve, the absolute half-point and relative half-point
occur when the force ratio reaches 1/2 and \((1+c)/2\), respectively.
The corresponding velocity ratios \(u_{1/2}\) and \(u_{(1+c)/2}\) serve
as useful reference points for analysis and design.

\subsubsection{Energy consumption}\label{energy-consumption}

The total energy consumption, \(E\), consists of the energy expended by
both the penetration force \(E_Q\) and the applied torque \(E_T\), as
detailed in Equation~\ref{eq-energy}.

\begin{equation}\phantomsection\label{eq-energy}{ E = E_{Q}+E_{T} = \int_{0}^{L} Q_{z}\mathrm{d}z + \int_{0}^{\Theta}T_{z} \mathrm{d}\theta  }\end{equation}

\(Q_z\) and \(T_z\) denote the total penetration force and torque at
depth \(z\), while \(L\) and \(\Theta\) refer to their corresponding
total penetration depth and rotational angle, respectively.
\(\mathrm{d}\theta\) represents the incremental rotational angle at
depth \(z\).

For a constant velocity ratio \(u\), the incremental rotational angle
can be related to the incremental depth,
\(\mathrm{d}\theta = 2u/D \mathrm{d}z\). Using this relationship, the
energy consumed by the torque can be calculated using
Equation~\ref{eq-torqueenergy}.

\begin{equation}\phantomsection\label{eq-torqueenergy}{ E_{T} = \int_{0}^{L} T_z \cdot \frac{2u}{D}\mathrm{d}z  }\end{equation}

Figure~\ref{fig-analysisexample} (b) demonstrates the process for
calculating the energy consumed by torque. First, the torque curve is
approximated using a sextic polynomial, shown as the dotted blue curve.
The energy is then estimated by integrating this polynomial function
using the Simpson's rule as implemented in the SciPy Python library
\citep{SCIPY20}. This integration corresponds to the shaded area, which
is subsequently scaled by the relevant \(2u/D\) values
(Equation~\ref{eq-torqueenergy}) for different rotational cases. The
energy consumed by the penetration force is calculated in a similar
manner, except that the energy is the shaded area under the force curve
itself (Equation~\ref{eq-energy}).

\section{Results}\label{results}

The following section presents the outcomes of rotational penetration
experiments conducted under various conditions, including shallow, deep,
and confining pressure conditions. Detailed analyses of penetration
force, torque, force reduction, and energy consumption are provided.

\subsection{Shallow Rotational Penetration
	Tests}\label{shallow-rotational-penetration-tests}

\subsubsection{Penetration Force and Torque under Shallow
	Conditions}\label{penetration-force-and-torque-under-shallow-conditions}

\begin{figure}

	\centering{

		\pandocbounded{\includegraphics[keepaspectratio]{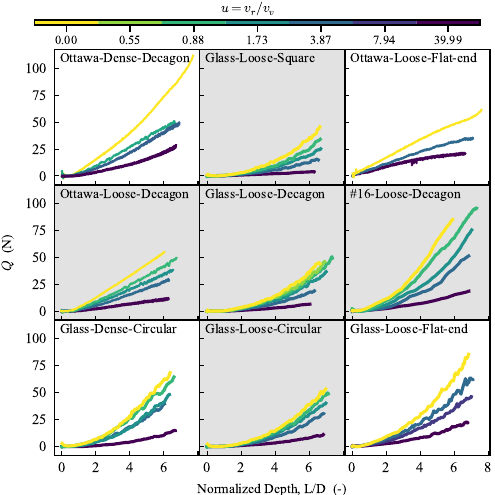}}

	}

	\caption{\label{fig-shallowforce}Penetration force under shallow
		conditions. (The colormap represents different velocity ratios, \(u\).
		This also applies to Figure~\ref{fig-shallowtorque}.)}

\end{figure}

\begin{figure}

	\centering{

		\pandocbounded{\includegraphics[keepaspectratio]{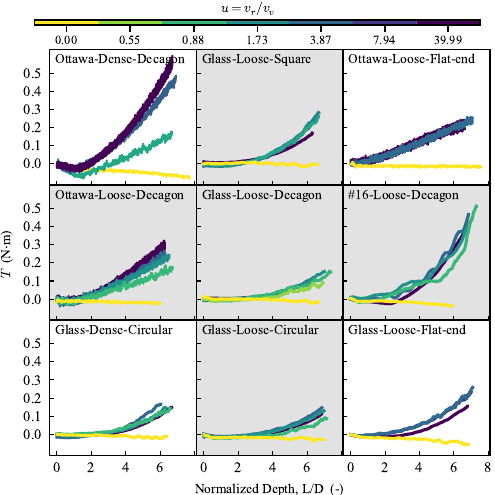}}

	}

	\caption{\label{fig-shallowtorque}Penetration torque under shallow
		conditions.}

\end{figure}

Figure~\ref{fig-shallowforce} and Figure~\ref{fig-shallowtorque}
illustrate examples of the penetration force (\(Q\)) and torque (\(T\))
under shallow conditions. Generally, \(Q\) increases with penetration
depth but decreases with \(u\). In contrast, \(T\) increases with both
variables. After the full immersion of the cone (\(0 < L \leq 0.87 D\)),
where \(Q\) increases slightly, a more rapid increase of \(Q\) occurs
during the shallow penetration stage (\(D < L \leq 8 D\)). The
penetration force curves exhibit upward concavity at deeper depths of
the shallow penetration stage, aligning with the ``initial penetration
phase'' proposed by \citep{PF02}. Both the decrease in \(Q\) and the
increase in \(T\) are not proportional to the increase in rotational
speed nor velocity ratio. Specifically, \(Q\) decreases slightly at
lower velocity ratios but rapidly at higher velocities. In contrast,
\(T\) increases significantly at lower velocity ratios but gently at
higher velocities. Theoretically, \(T\) in control cases (\(u = 0\))
should always be zero since there is no rotation. However, slightly
increases of \(T\) with penetration depth were observed in these cases,
likely due to multiple connections between the rotary motor and
penetrator causing imperfect alignment of the penetration system.

\subsubsection{Penetration Force Reduction under Shallow
	Conditions}\label{penetration-force-reduction-under-shallow-conditions}

\begin{figure*}

	\centering{

		\pandocbounded{\includegraphics[keepaspectratio]{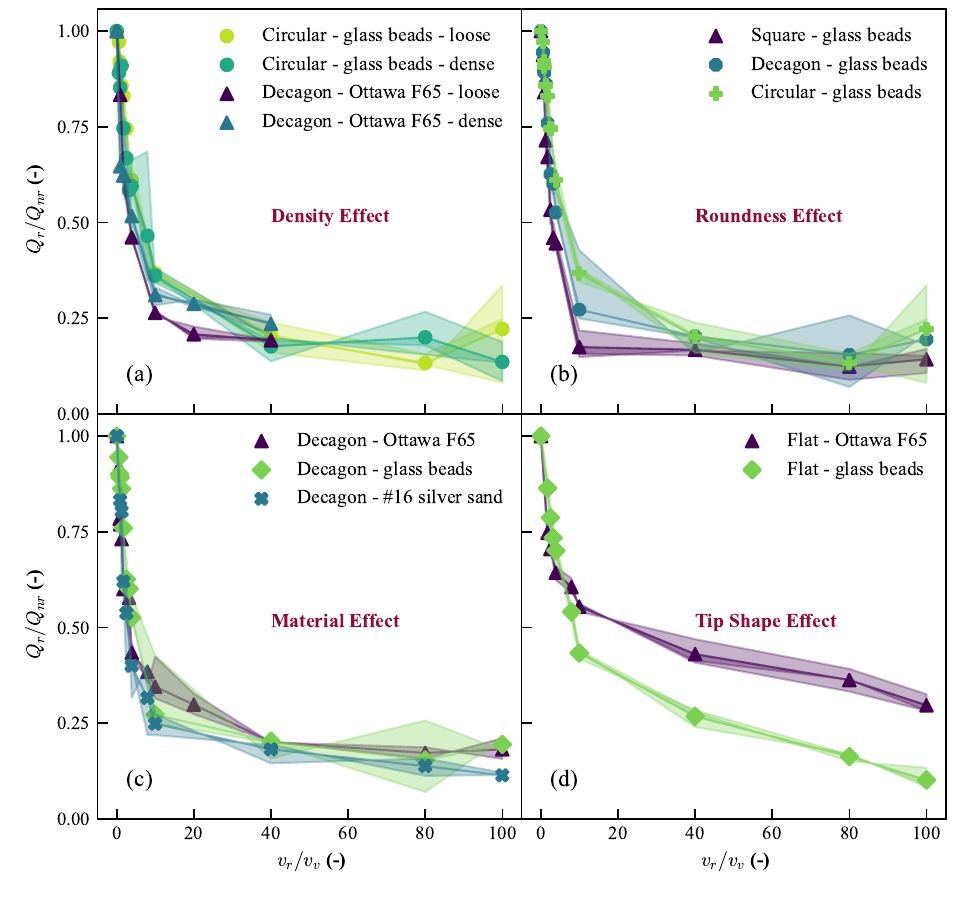}}

	}

	\caption{\label{fig-multieffect}Effects on the reduction of penetration
		force under shallow conditions: (a) influence of density; (b) impact of
		penetrator roundness; (c) effect of material type; (d) effect of tip
		shape and material type. The curves represent the average values across
		trials, with shaded areas indicating the standard deviations.}

\end{figure*}

The reductions in penetration force for shallow rotational penetration
tests, shown in Figure~\ref{fig-multieffect}, are analyzed based on
various factors, including relative density, penetrator roundness,
material types and tip shape. In all but the flat-end cases,
\(Q_{r}/Q_{nr}\) decreases sharply then levels off to about 0.2 as \(u\)
increases. The fitting parameters and absolute and relative half-point
velocity ratios for all testing cases, according to
Equation~\ref{eq-fit} and Figure~\ref{fig-analysisexample} (c) are
summarized in Table~\ref{tbl-fittingparams}. For clarity, the range
\(u = 0\) to \(u_{1/2}\) is termed ``low velocity ratios,'' and values
beyond \(u_{1/2}\) as ``high velocity ratios.'' The parameters \(k\) and
\(c\) indicate the reduction rate and the ultimate force ratio,
respectively. Higher \(k\) values imply faster reduction and thus lower
half-point velocity ratios, while lower \(c\) values indicate a smaller
ultimate force ratio.

Figure~\ref{fig-multieffect} (a-c) show that the relative density and
type of the granular media, as well as penetrator roundness all have
limited effects on penetration force reduction, especially at high
velocity ratios (similar \(c\) values in Table~\ref{tbl-fittingparams}).
Relative density has more impact on Ottawa F65 sand (higher \(\Delta k\)
and \(\Delta c\)) than glass beads; for both materials, reduction is
faster in denser materials (higher \(k\)); the ultimate force ratio
increase with density for both materials, which is more obvious in
Ottawa F65 sand. Lower penetrator roundness leads to faster and greater
reduction (higher \(k\) and lower \(c\)). The penetration force ratio
reduces more slowly with \(u\) in glass beads compared to Ottawa F65
sand and \#16 silver sand, but reaches similar ultimate force ratio for
in all materials (Figure~\ref{fig-multieffect} (c-d),
Table~\ref{tbl-fittingparams}). The differences between the glass beads
and sands is more significant when the penetrator has a flat end instead
of a cone tip (Figure~\ref{fig-multieffect} (d)). Among all the factors
investigated, the tip shape has the most significant effect on the
reduction of penetration force (comparing Figure~\ref{fig-multieffect}
(c) with (d)). Flat-ended penetrators exhibit much less force reductions
compared to coned penetrators, and the reductions do not stabilize even
at the highest velocity ratio studied in this study (\(u = 100\)).

\begin{table*}

	\caption{\label{tbl-fittingparams}Fitting parameters and half-life velocity ratio for all cases }

	\centering{

		\centering

		\begin{tabular}{l|l|l|l|l|l}
			\hline
			Groups                     & $c$  & $k$  & $R^2$ & $u_{1/2}$ & $u_{(1+c)/2}$ \\
			\hline
			Circular-loose-glass beads & 0.10 & 0.17 & 0.99  & 7.25      & 5.76          \\
			Decagon-loose-Ottawa F65   & 0.11 & 0.38 & 0.99  & 3.35      & 2.61          \\
			Circular-dense-glass beads & 0.11 & 0.22 & 0.98  & 5.82      & 4.53          \\
			Decagon-dense-Ottawa F65   & 0.23 & 0.63 & 0.97  & 2.89      & 1.58          \\
			Decagon-glass beads        & 0.11 & 0.25 & 0.98  & 5.19      & 4.04          \\
			Decagon-\#16 silver sand   & 0.09 & 0.38 & 0.98  & 3.21      & 2.69          \\
			Square-glass beads         & 0.08 & 0.37 & 0.97  & 3.26      & 2.66          \\
			Flat-glass beads           & 0.07 & 0.12 & 0.99  & 9.26      & 8.04          \\
			Flat-Ottawa F65            & 0.34 & 0.27 & 0.96  & 11.39     & 3.69          \\
			Deep-PLA-Cone              & 0.12 & 0.31 & 0.99  & 4.2       & 3.20          \\
			Deep-PLA-Flat              & 0.34 & 0.29 & 0.99  & 11.0      & 3.50          \\
			Deep-Steel-Flat            & 0.48 & 0.28 & 0.98  & 90.96     & 3.53          \\
			Confining-2.66 kPa         & 0.16 & 0.42 & 0.99  & 4.83      & 2.41          \\
			Confining-5.77 kPa         & 0.19 & 0.34 & 0.99  & 4.20      & 2.96          \\
			\hline
		\end{tabular}

	}

\end{table*}%

\subsubsection{Energy Consumptions under Shallow
	Conditions}\label{energy-consumptions-under-shallow-conditions}

\begin{figure*}

	\centering{

		\pandocbounded{\includegraphics[keepaspectratio]{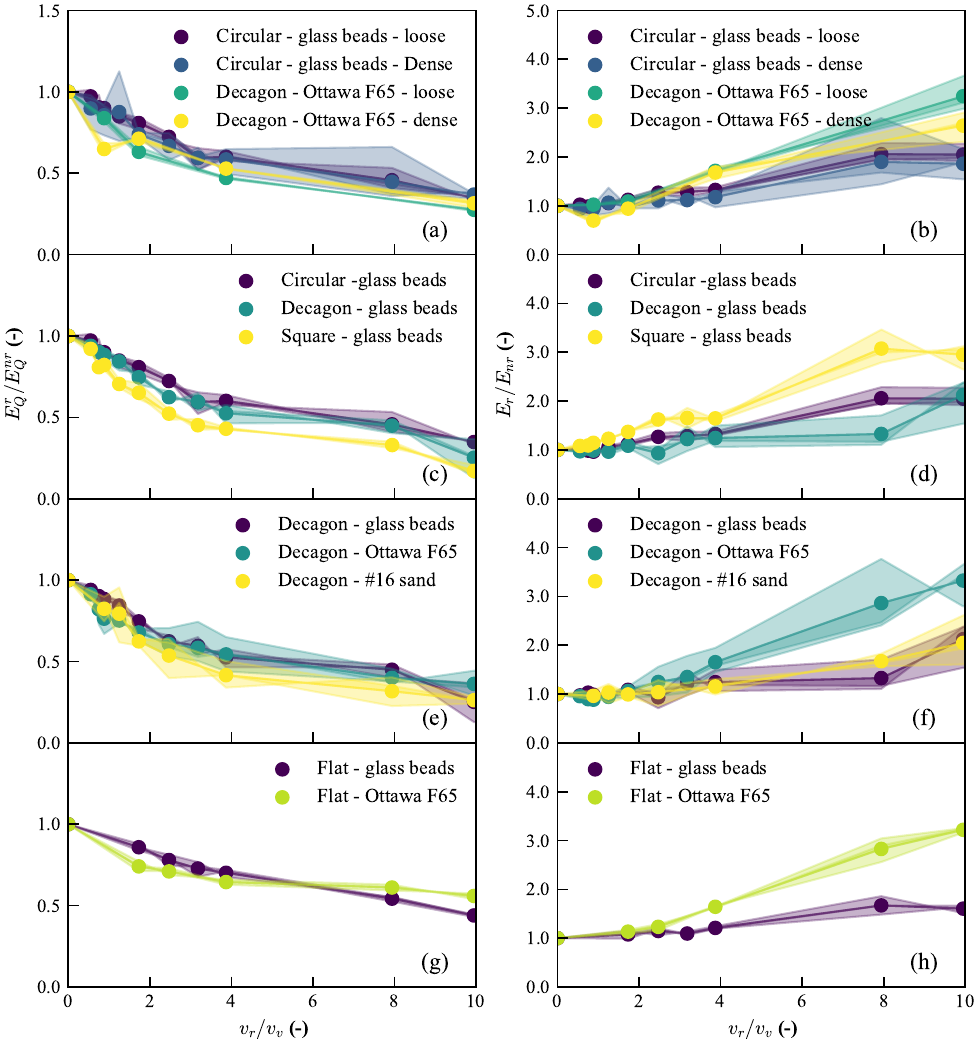}}

	}

	\caption{\label{fig-shallowenergy}Energy consumption under shallow
	conditions: (a) and (b) for density effect; (c) and (d) for penetrator
	roundness effect; (e) and (f) for material effect; (g) and (h) for
	flat-end comparison. (The left column shows the force-based energy
	ratio, \(E_{Q}^{r}/E_{Q}^{nr}\), while the right column shows the total
	energy ratio, \(E_{r}/E_{nr}\).)}

\end{figure*}

The force-based energy ratio, defined as the ratio of energies consumed
by force in rotational and non-rotational penetrations,
\(E_{Q}^{r}/E_{Q}^{nr}\), and the total energy ratio, \(E_{r}/E_{nr}\),
under shallow conditions are illustrated in
Figure~\ref{fig-shallowenergy}. The energy plots are limited to
\(u \leq 10\) to highlight the reduction before the half-points.
Generally, force-based energy ratios decreases with \(u\), while total
energy ratios increases across all cases due to rotation. The reduction
of the force-based energy ratio is greater for decagon-Ottawa cases
compared to circular-glass cases across various relative densities,
whereas the total energy ratio is greater for circular-glass cases, as
shown in Figure~\ref{fig-shallowenergy} (a) and (b). Decreasing the
roundness of the penetrator slightly decreases force-based energy ratio
at the same velocity ratio (Figure~\ref{fig-shallowenergy} (c)), but
increases the total energy ratio (Figure~\ref{fig-shallowenergy} (d)).
As shown in Figure~\ref{fig-shallowenergy} (e) and (f), with the same
decagon penetrator, the force energy ratio is comparable between glass
beads and Ottawa F65 materials but is slightly lower for \#16 silver
sand cases; in contrast, the total energy ratio for the Ottawa F65 sand
case is higher than for glass beads and \#16 silver sand cases. Similar
trends were observed with flat-end penetrator
(Figure~\ref{fig-shallowenergy} (g) and (h)).

A consistent finding across all cases is that at relatively low velocity
ratios (lower than 3), the force-induced energy reduces rapidly while
the total energy remains relatively unchanged. This indicates that a
slight rotation can cause considerable reduction in penetration force
without additional cost.

\subsection{Deep Rotational Penetration
	Tests}\label{deep-rotational-penetration-tests}

Deep rotational penetration tests were conducted in loose Ottawa F65
sand using three types of penetrators, as shown in
Figure~\ref{fig-penetrator} (f), (g), and (h). These tests also aimed to
further investigate the effects of penetrator-material interface
friction and tip shape on penetration force reduction.

\subsubsection{Penetration Force and Torque under Deep
	Conditions}\label{penetration-force-and-torque-under-deep-conditions}

\begin{figure*}

	\centering{

		\pandocbounded{\includegraphics[keepaspectratio]{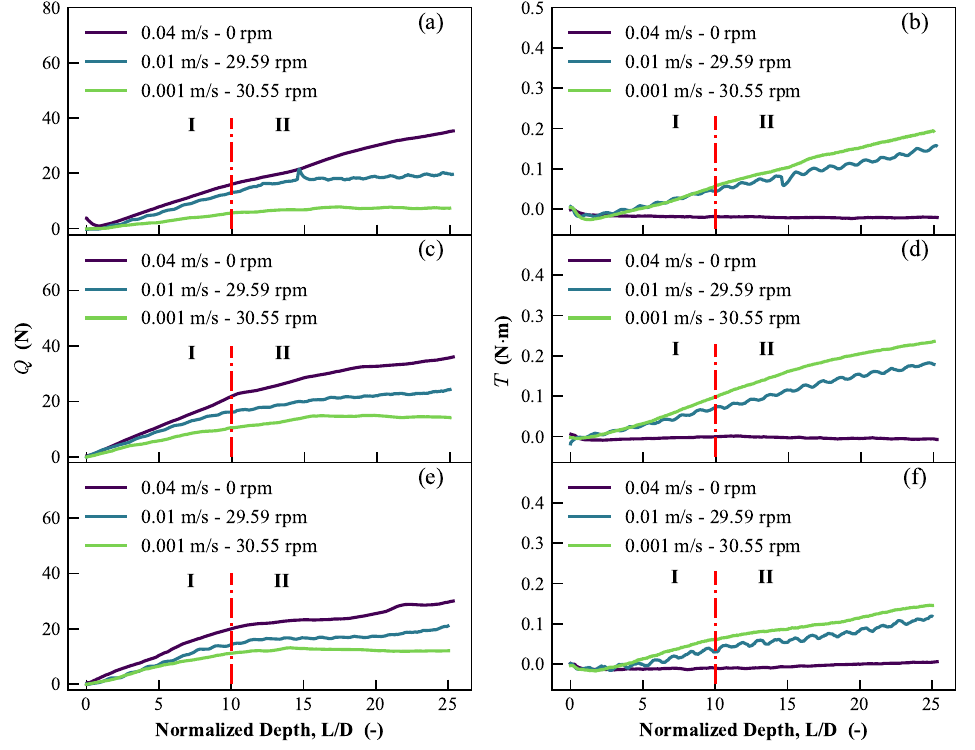}}

	}

	\caption{\label{fig-deeppenetration}\(Q\) and \(T\): (a) and (b) for
		PLA-cone penetrator; (c) and (d) for PLA-flat penetrator; (e) and (f)
		for Steel-flat penetrator.}

\end{figure*}

The penetration force and torque data from nine deep rotational
penetration cases are presented in Figure~\ref{fig-deeppenetration}.
Consistent with the results from shallow tests, the penetration force,
\(Q\), decreases, while the torque, \(T\), increases with \(u\).
However, for \(Q\), two distinct stages are evident: Stage I (shallow
penetration) with a penetration depth of less than a critical depth,
\(L_{crit}\), \(Q\) increases relatively quickly, followed by Stage II
(deep penetration) with more gradual increase, as shown in
Figure~\ref{fig-deeppenetration} (a), (c), and (e). The critical depth
\(L_{crit}\) differs among cases, but is usually around 10 \(D\). The
difference in the rate of increase between the two stages is more
pronounced in the flat-end penetrator cases. Similarly, \(T\) followed a
comparable two-stage trend, although the distinction between the two
stages is less pronounced compared to \(Q\).

\subsubsection{Penetration Force Reduction and Energy Consumption under
	Deep
	Conditions}\label{penetration-force-reduction-and-energy-consumption-under-deep-conditions}

\begin{figure*}

	\centering{

		\pandocbounded{\includegraphics[keepaspectratio]{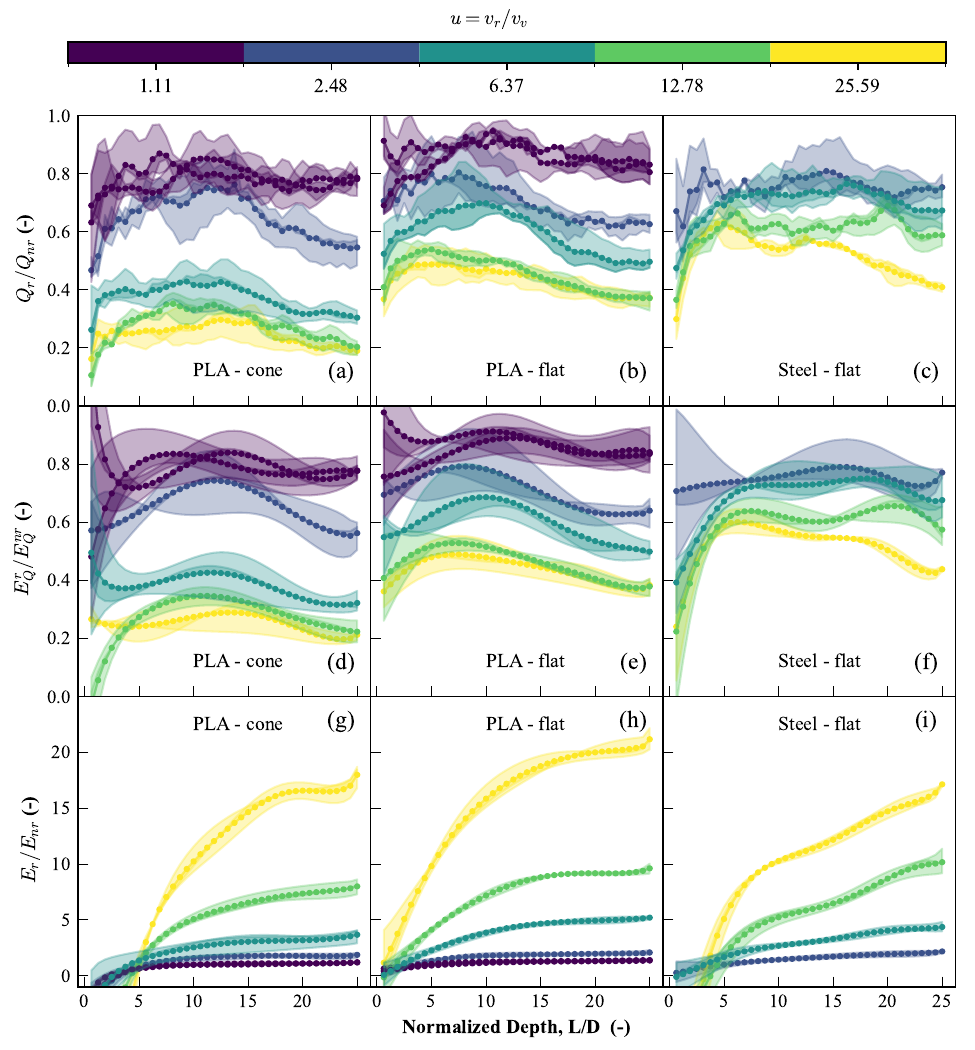}}

	}

	\caption{\label{fig-deepreductiondepth}Variations of \(Q_{r}/Q_{nr}\),
	\(E_{Q}^{r}/E_{Q}^{nr}\), \(E_{r}/E_{nr}\) with penetration depth: (a),
	(d), (g) for PLA-cone cases; (b), (e), (h) for PLA-flat cases; (c), (f),
	(i) for Steel-flat cases.}

\end{figure*}

Figure~\ref{fig-deepreductiondepth} presents the force ratios,
force-based energy ratios and total energy ratios along the penetration
depth at different velocity ratios. As the penetration depth increase,
the force ratio \(Q_{r}/Q_{nr}\) initially increases and then decreases
for all velocity ratios. \(Q_{r}/Q_{nr}\) is comparable at shallower
depths across all cases with \(u = 2.48\). However, at deeper depths,
\(Q_{r}/Q_{nr}\) sharply decreases for PLA-cone and PLA-flat cases,
whereas the decrease is more gradual for Steel-flat case. Additionally,
the inflection point, which marks the transition from increasing to
decreasing trend, tends to shift to a shallower depth as \(u\)
increases. Such inflections in \(Q_{r}/Q_{nr}\) along depth is
hypothesized to be associated with the critical depth \(L_{crit}\) and
the different failure modes of shallow versus deep penetrations.

The fitting parameters of the force reduction curves (\(Q_{r}/Q_{nr}\)
vs \(u\)) for the deep test groups are shown in
Table~\ref{tbl-fittingparams}. Comparing PLA and steel penetrators,
increased interface roughness slightly increases the reduction rate
(\(k\): 0.28 vs 0.29) and considerably lowers the ultimate force ratio
(\(c\): 0.48 vs 0.34). For PLA penetrators with different tip shapes, a
coned tip results in a minor increase in the reduction rate (\(k\): 0.29
vs 0.31) and a substantial decrease in the ultimate force ratio (\(c\):
0.34 vs 0.12). In the comparison between the Deep-Steel-Flat case with
the Flat-Ottawa F65 case, which differ only in penetration depth, deeper
penetration leads to slight increase in the reduction rate (\(k\): 0.27
to 0.28) but causes a considerable increase in the ultimate force ratio
(\(c\): 0.34 to 0.48). These results confirm the effect of the cone tip
on force reduction observed in the shallow cases and further demonstrate
that a rougher interface enhances force reduction, while deeper
penetration inhibits it.

In terms of energy, while the trends of force-based energy ratios
generally mirror those of force ratios, the total energy ratios show a
more straightforward pattern. The total energy ratio increases with all
the factors, including interface roughness, penetration depth, tip
bluntness and velocity ratio.

\subsection{Rotational Penetration Tests under Confining
	Pressures}\label{rotational-penetration-tests-under-confining-pressures}

Due to the load capacity limitations of the robotic arm, the confining
pressures applied in the final test groups were relatively low, at 2.66
kPa and 5.77 kPa. Nonetheless, the results still demonstrated clear and
consistent trends.

\subsubsection{Penetration Force and Torque under Confining Pressure
	Conditions}\label{penetration-force-and-torque-under-confining-pressure-conditions}

\begin{figure*}

	\centering{

		\pandocbounded{\includegraphics[keepaspectratio]{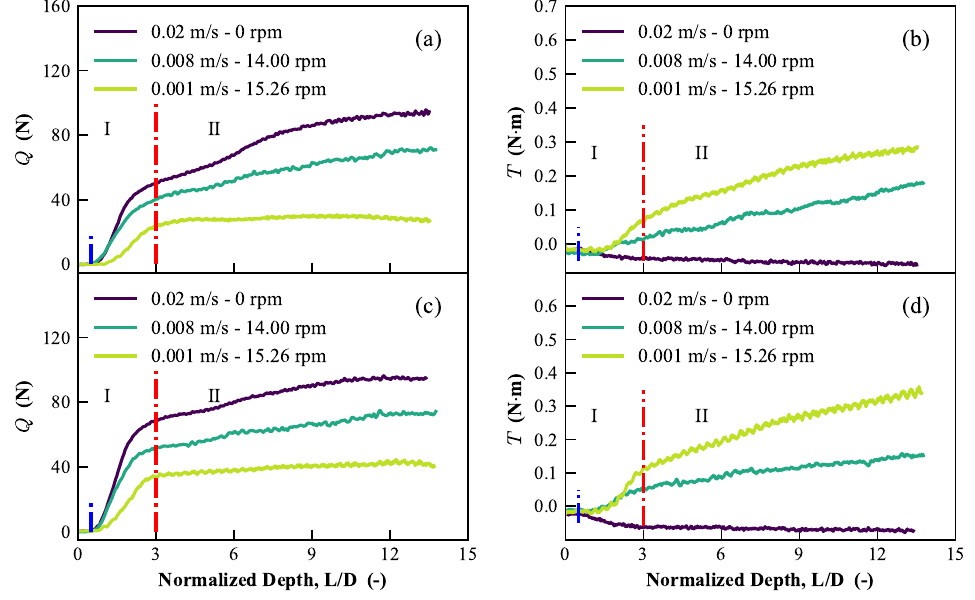}}

	}

	\caption{\label{fig-confiningpenetration}Penetration force and torque
		under with confining pressures: 2.66 kPa for (a) and (b); and 5.77 kPa
		for (c) and (d).}

\end{figure*}

The force and torque data for six representative tests with confining
pressures are presented in Figure~\ref{fig-confiningpenetration}. As
observed under deep conditions, both \(Q\) and \(T\) displays two
distinct stages. Note that the initial penetration depth of \(0.5D\),
indicated by the blue dashed line in
Figure~\ref{fig-confiningpenetration}, corresponds to the thickness of
the 3D-printed plate used to apply the pressure
(Figure~\ref{fig-container}).

Compared to the deep condition, Stage I exhibits a more rapid increase
in \(Q\) and the critical depth, \(L_{crit}\), shifts to a shallower
depth. Under the same velocity ratio \(u\), the increasing rate of \(Q\)
in Stage I is higher under higher confining pressure. Additionally, at
the final depth, the force \(Q\) for the non-rotational cases remain
similar under different confining pressures. For both pressure
conditions, the torque \(T\) in the non-rotational case exhibits a
negative value during penetration. However, this torque is negligible in
magnitude compared to the rotational cases, and is likely a result of
slight vertical misalignment in the setup.

\subsubsection{Penetration Force Reduction and Energy Consumption under
	Confining Pressure
	Conditions}\label{penetration-force-reduction-and-energy-consumption-under-confining-pressure-conditions}

\begin{figure*}

	\centering{

		\pandocbounded{\includegraphics[keepaspectratio]{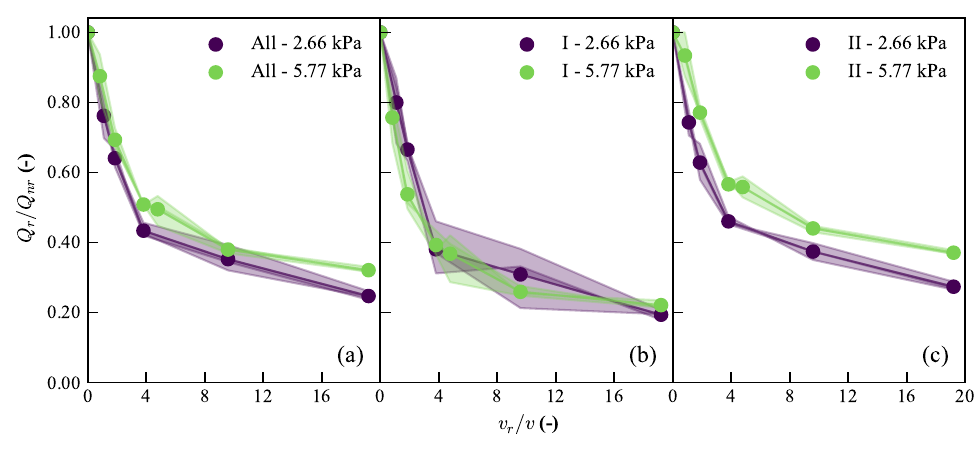}}

	}

	\caption{\label{fig-confiningreduction}Force ratio \(Q_{r}/Q_{nr}\)
	under confining pressures: (a) overall reduction; (b) reduction in
	Stage; (c) reduction in Stage II.}

\end{figure*}

Figure~\ref{fig-confiningreduction} demonstrates the penetration force
reduction under confining pressure conditions. Two stages were
considered separately to calculate \(Q_{r}/Q_{nr}\), corresponding to
the two distinct phases of the penetration force in
Figure~\ref{fig-confiningpenetration}. For each confining pressure, the
force ratio \(Q_{r}/Q_{nr}\) in Stage I decreases more rapidly than in
Stage II at the same \(u\). In Stage I, \(Q_{r}/Q_{nr}\) is similar
across all velocity ratios for both confining pressures. In contrast, it
increases with confining pressure in Stage II. For example, at
\(u = 19.2\), \(Q_{r}/Q_{nr}\) ratios are 0.19 and 0.22 for Stage I, but
0.27 and 0.37 for Stage II, leading to overall values of 0.25 and 0.32,
respectively. Overall, increasing confining pressure decreases reduction
rate (\(k\): 0.42 to 0.34) and increases the ultimate force ratio
(\(c\): 0.16 to 0.19), as shown in Table~\ref{tbl-fittingparams}.

\begin{figure*}

	\centering{

		\pandocbounded{\includegraphics[keepaspectratio]{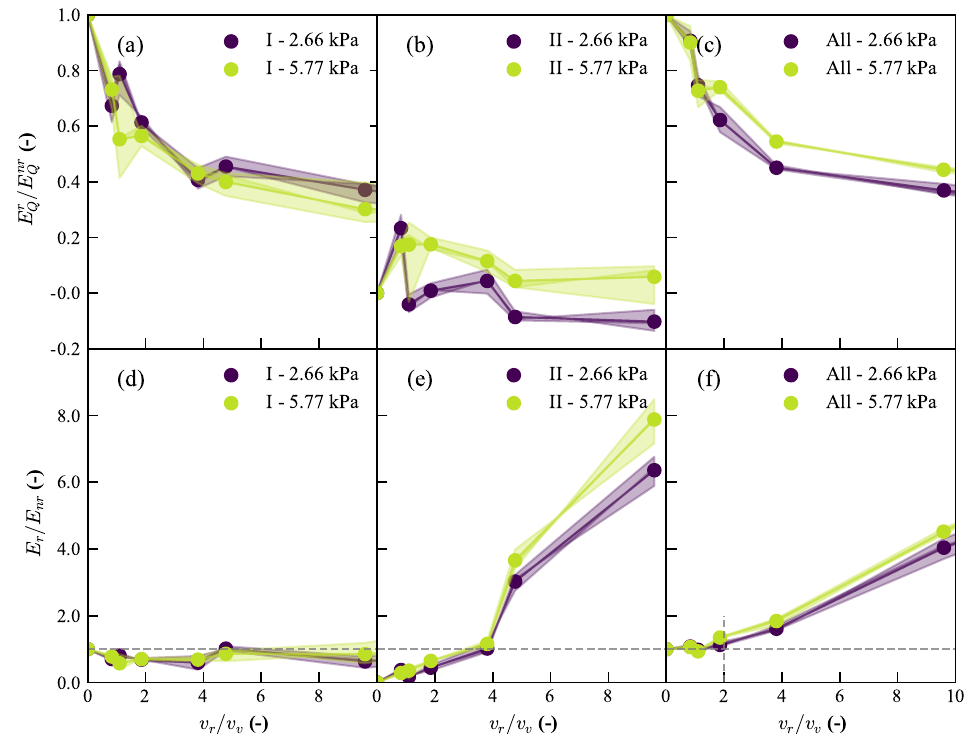}}

	}

	\caption{\label{fig-confiningenergy}Force-based energy ratio
	\(E_{Q}^{r}/E_{Q}^{nr}\), and the total energy ratio, \(E_{r}/E_{nr}\),
	under confining pressure conditions: (a) and (d) for Stage I; (b) and
	(e) for Stage II; (c) and (f) for whole process.}

\end{figure*}

The force-based and total energy ratios under confining pressure
conditions are presented in Figure~\ref{fig-confiningenergy}, also with
two stages. In Stage I, the force-based energy ratio,
\(E_{Q}^{r}/E_{Q}^{nr}\), are nearly identical for both pressure
conditions, decreasing to 0.3 at \(u = 10\). Similarly, the total energy
ratio, \(E_{r}/E_{nr}\), are comparable between the two pressure
conditions, but consistently remaining around 1 as \(u\) increases from
0 to 10. In Stage II, the trends are different: \(E_{Q}^{r}/E_{Q}^{nr}\)
for the higher confining pressure case are slightly higher than those
for lower pressure case, but remain below 0.2 throughout the process. In
contrast, \(E_{r}/E_{nr}\) increases with velocity ratio, and it is
higher for the higher pressure case when \(u > 4\). Overall,
Figure~\ref{fig-confiningenergy} (c) and (f) indicate that both
\(E_{Q}^{r}/E_{Q}^{nr}\) and \(E_{r}/E_{nr}\) increase marginally with
confining pressure under same velocity ratio.

\section{Discussions}\label{discussions}

\subsection{Synthesis of Trends on Rotation-induced Reduction of
	Penetration
	Force}\label{synthesis-of-trends-on-rotation-induced-reduction-of-penetration-force}

\begin{figure*}

	\centering{

		\pandocbounded{\includegraphics[keepaspectratio]{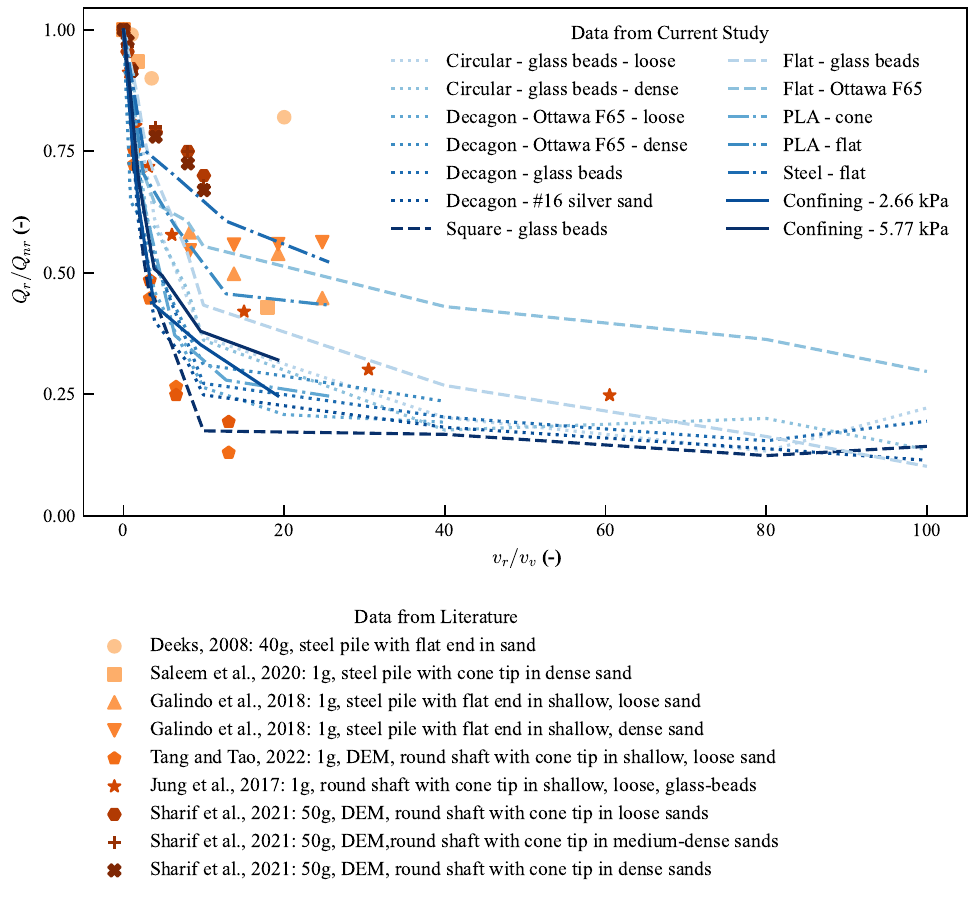}}

	}

	\caption{\label{fig-synthesis}Synthesis of normalized rotational
		penetration force data from current study and the literature}

\end{figure*}

Data from current and previous studies on rotational penetration force
reduction are summarized in Figure~\ref{fig-synthesis}. The data
consistently support an decrease in force ratio \(Q_{r}/Q_{nr}\) as
\(u\) increases. However, the force reduction rates and ultimate force
ratios vary significantly across the datasets. With the broader ranges
of \(u\) considered in this study, the decline of force ratio is further
confirmed to follow an approximately hyperbolic trend.

Theoretical expressions for the rotational penetration force have been
proposed in several studies \citep{BMW97, SBC+21, YZW+24}. These
derivations typically rely on a common set of assumptions, such as
uniform soil properties at the cone scale and the invariance of normal
stress on the cone surface under rotation. However, our previous
numerical simulations demonstrated a significant reduction in normal
stress due to rotation \citep{TT22b}. By relaxing this assumption and
employing a vector-based derivation, we derived expressions for the
total penetration resistance of a rotating cone (Equation~\ref{eq-qcr})
and a non-rotating cone (Equation~\ref{eq-qcnr}), as well as for the
penetration resistance on rotating and non-rotating shafts
(Equation~\ref{eq-fz}). We then obtained expressions for the total
penetration forces for rotating and non-rotating configurations
(Equation~\ref{eq-Qr}, Equation~\ref{eq-Qnr}) and their ratio
(Equation~\ref{eq-analytical-ratio}). The results are consistent with
those in \citet{YZW+24}.

Equation~\ref{eq-analytical-ratio} shows that the penetration resistance
on both the rotating cone and shaft follows hyperbolic functions of the
relative velocity ratio \(u\) (Equation~\ref{eq-phi-cu},
Equation~\ref{eq-phi-su}). Although the sum of two hyperbolic functions
does not inherently result in another hyperbolic function, the total
penetration force can be reasonably approximated as one. This
approximation is valid because, in sand, the shaft penetration force is
typically much smaller compared to the cone penetration force,
especially at shallow penetration depths. This theoretical analysis
clarifies that a hyperbolic decay is more appropriate and justified than
an exponential decay \citep{JCKK17c}, which aligns with the nearly
hyperbolic trend observed in the force ratio data from this study. When
\(u\) approaches infinity, the contributions of friction on the total
rotational penetration force, both on the cone and shaft, approach zero,
leading to the ultimate force ratio as shown in
Equation~\ref{eq-ratio-limit}.

Flat-ended penetrators can be conceptualized as cones with a half-angle,
\(\alpha\), of 90 degrees. As demonstrated by
Equation~\ref{eq-ratio-limit}, the relative force ratio increases with
increasing \(\alpha\). This observation is consistent with result in
shallow penetration tests (comparing Figure~\ref{fig-multieffect} (c)
with (d)) and provides a partial explanation for the differences
observed between the deep penetration behaviors of the PLA-cone and
PLA-flat cases, as illustrated in Figure~\ref{fig-deeppenetration} (a)
and (b), respectively. Furthermore, Equation~\ref{eq-ratio-limit}
indicates that the relative force ratio decreases as the interface
friction coefficient, \(\mu\), increases. This trend aligns with the
experimental results comparing the Steel-flat and PLA-flat cases, as
shown in Figure~\ref{fig-deeppenetration} (b) and (c). The influence of
tip shape and interface friction coefficient also explains the
relatively high ultimate force ratio reported in \citet{GDB18}, which
involved shallow penetration of a steel flat-ended pile in sands
(Figure~\ref{fig-synthesis}).

As depth and confining pressure the total non-rotational penetration
force (\(Q_{nr}\)) is expected to increase. For the ultimate force ratio
(\(Q_{r}/Q_{nr}\)) to remain unchanged under these conditions, the
rotational cone normal force (\(\sigma_c^{r}A\)) must change
proportionally with \(Q_{nr}\) (Equation~\ref{eq-ratio-limit}). However,
the available data suggests disproportional changes
(Figure~\ref{fig-synthesis}). For instance, increases in depth
(Figure~\ref{fig-deeppenetration} (a)-(c)) and confining pressure
(Figure~\ref{fig-confiningreduction}) have been shown to lead to higher
ultimate values of \(Q_{r}/Q_{nr}\).

Although the direct effect of gravity is not examined in this study, an
elevated gravitational field is expected to increase the stress level in
sand, resembling deep penetration under higher confining pressures.
Consequently, higher \(g\)-levels are anticipated to reduce the
effectiveness of rotation-induced penetration resistance reduction.
Overall, all the inhibiting factors, including higher \(g\) level, lower
interface friction and a flat-tip design, explains the observed low
reduction rate and high ultimate force ratio reported in \citet{Dee08}
(Figure~\ref{fig-synthesis}), where penetration was conducted using a
flat-ended steel penetrator in sand under 40 \(g\).

The effect of relative density on \(Q_{r}/Q_{nr}\) remains inconclusive
(Figure~\ref{fig-synthesis}). Data from this study and \citet{GDB18}
indicate that rotational penetration of a closed-ended penetrator in
denser sands under 1\(g\) conditions slightly increases
\(Q_{r}/Q_{nr}\). In contrast, numerical simulations by \citet{SBC+21}
under 50\(g\) conditions show the opposite trend, though the differences
across cases are small. In the latter case, it is plausible that the
elevated gravitational field significantly increased the confining
pressure, overshadowing the influence relative density.

One of the more consistent trend observed in this study is the effect of
penetrator cross-section shape. Figure~\ref{fig-multieffect} (b)
demonstrates that as penetrator roundness decreases, \(Q_{r}/Q_{nr}\)
decreases, especially when \(u<20\). On the other hand, influences of
granular material characteristics (e.g., particle shape and size
distribution) on \(Q_{r}/Q_{nr}\) are less clear. While these influences
are limited for conical penetrators (Figure~\ref{fig-multieffect} (c)),
these factors become significantly more pronounced for flat-ended
penetrators.

In summary, rotation induces reductions in both contributions from
friction-based shear stress ans well as contact-based cone normal stress
\(\sigma_c^{r}\), but rotation's relative effect depends on intrinsic
properties of the granular material and the penetrator, as well as
external factors such as gravitational field and confining pressure. It
is observed that these critical factors overlap strongly with those
governing interface dilatancy. This parallelism suggests that
micromechanical theories of dilatancy (e.g., particle rearrangement,
shear-induced volume changes) may provide a unified pathway to model
rotational effects.

\subsection{Energy Implications}\label{energy-implications}

The energy consumption during rotational penetration tests under
shallow, deep, and confining pressure conditions is illustrated in
Figure~\ref{fig-shallowenergy}, Figure~\ref{fig-deepreductiondepth}, and
Figure~\ref{fig-confiningenergy}. In general, the relative energy
associated with force (\(E_{Q}^{r}/E_{Q}^{nr}\)) decreases sharply at
first and then stabilizes as \(u\) increases. Conversely, the total
relative energy (\(E_{r}/E_{nr}\)) increases with \(u\), starting
gradually but accelerating at higher values, primarily due to the
growing energy consumption from torque.

\begin{figure*}

	\centering{

		\pandocbounded{\includegraphics[keepaspectratio]{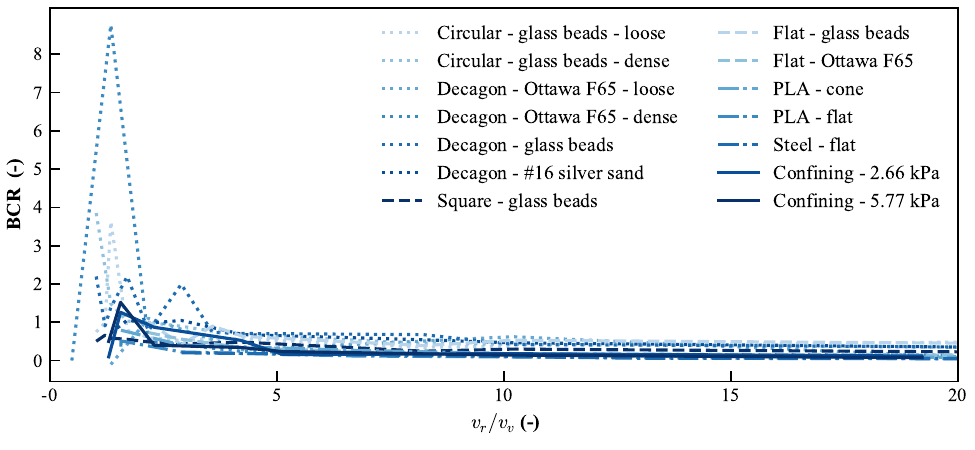}}

	}

	\caption{\label{fig-bcr}Average benefit-cost ratio for each test group.}

\end{figure*}

The benefit-cost ratio (\(\text{BCR}\)) is defined as the ratio of the
benefit---reduction in energy associated with force---to the cost,
represented by the energy consumed by torque (Equation~\ref{eq-bcr}).
The \(\text{BCR}\) for all cases is summarized in Figure~\ref{fig-bcr}.

\begin{equation}\phantomsection\label{eq-bcr}{
	\text{BCR} = \frac{\Delta E_Q}{\Delta E_T} = \frac{E_Q^{nr} - E_Q^{r}}{E_{r} - E_{nr}}
	}\end{equation}

Results show that \(\text{BCR}\) falls below 1 at high \(u\) values,
indicating energy inefficiency at greater rotational speeds. The maximum
\(\text{BCR}\) is typically observed at relatively low \(u\) values
(approximately 0--3), depending on the test conditions. These findings
align with simulation results from \citeyearpar{CM}, which identified an
optimal rotational velocity (\(u < 1\)) that minimizes both penetration
force and total energy. However, it should be noted that the penetrator
in the referenced study features a conical tip bent from the vertical to
an inclined orientation, which differs from the configurations analyzed
here.

Our prior work demonstrated that dual-anchor burrowing can be achieved
in a minimalistic robot through coordinated rotation and
extension/contraction \citep{TT22b, TZTa, ZHT23, ZT23}. For such
self-burrowing robots, two critical design challenges emerge: reducing
penetration resistance and minimizing energy consumption. This study
reveals a key trade-off: higher velocity ratios lower penetration forces
but demand greater energy input, whereas lower ratios conserve energy at
the expense of increased burrowing force. To optimize performance,
future designs should integrate dynamic locomotion strategies, including
controlled rotation, to balance these competing priorities.

\subsection{Limitations and Future
	Work}\label{limitations-and-future-work}

The holistic rotational penetration tests in this study, while
illuminating key relationships between force reduction and its
influencing factors, are subject to several constraints. First, the
robotic arm's force capacity (limited to 150 N) restricted achievable
penetration depths, particularly under higher confining pressure, and
precluded systematic testing of penetrators with larger diameters.
Second, the applied confining pressure levels were orders of magnitude
lower than centrifuge-simulated conditions \citep{SBC+21}, potentially
underestimating scaling effects on force reduction. Third, while grain
shape is known to influence tip resistance \citep{LL}, the study did not
isolate the effects of grain size and shape across materials, limiting
mechanistic insights into their individual contributions. Third,
critical factors such as particle breakage (relevant in high-stress
environments) and saturation (critical for submarine or geotechnical
applications) were not modeled. Finally, the exclusive measurement of
total penetration forces and torques in this study precludes isolation
of rotational effects on tip resistance (\(Q_c\)) versus shaft
resistance (\(Q_s\)), limiting further mechanistic analysis of their
individual contributions.

Future work should prioritize two interconnected objectives: (1)
addressing the methodological limitations identified in this study, and
(2) establishing a unified mechanistic framework (e.g., interface
dilatancy) to explain how intrinsic and external factors govern
rotation-induced reductions in cone normal resistance.

\section{Conclusions}\label{conclusions}

This study has systematically investigated the effects of rotational
motion on penetration resistance and energy consumption in granular
media under shallow, deep, and low confining-pressure conditions. A
broad range of tangential-to-axial velocity ratios (up to
\(u \approx 100\)) was considered using multiple penetrator geometries,
interface friction levels, penetration depths, and particle types. Based
on the results, several key conclusions can be drawn:

\begin{itemize}
	\item
	      Hyperbolic decay of penetration force. Across all conditions, the
	      rotational penetration force \(Q_{r}\) decreases approximately in a
	      hyperbolic manner with increasing velocity ratio \(u\). At low
	      velocity ratios (e.g., \(u < 3\)), even modest rotation achieves
	      meaningful reductions in penetration force. Beyond a certain threshold
	      (the ``half-point'' velocity ratio), further increases in \(u\) offer
	      diminishing returns, with force ratios approaching an ultimate limit.
	\item
	      Effects of penetrator geometry and interface friction. Penetrators
	      with conical tips achieve greater force reductions than flat-ended
	      designs, underscoring the importance of apex angle. Cross-sectional
	      roundness also plays a significant role, particularly at lower
	      velocity ratios: less rounded geometries lead to faster and more force
	      reductions. In addition, rougher interfaces also facilitates
	      rotation-induced reductions of penetration resistance.
	\item
	      Effects of confining pressure and penetration depth. Higher confining
	      pressure and deeper penetration generally increase the overall
	      resistance and raise the ultimate force ratio, leading to reduced
	      effectiveness of rotational penetration. These increases are mainly
	      attributed to the deep penetration stage, which exhibits a different
	      failure mode compared to shallow penetration. Nevertheless, the core
	      hyperbolic trend remains consistent.
	\item
	      Limited impact of material type and relative density. Within the
	      ranges tested, the granular material (sands and glass beads) and its
	      relative density exert only moderate influences on force reduction,
	      especially at higher \(u\).
	\item
	      Energy trade-offs and design implications. While rotation consistently
	      lowers force-based energy consumption, it substantially raises
	      torque-related energy demands at higher \(u\). Consequently, total
	      energy consumption increases at large velocity ratios, even though the
	      net penetration force decreases. This trade-off suggests that
	      incorporating rotation at moderate \(u\) (\textless{} 3) may optimize
	      the balance between penetration force reduction and overall energy
	      cost, particularly relevant for self-burrowing robots that must manage
	      limited power and payload capacities.
\end{itemize}

In sum, this study provides a rich dataset and comprehensive analysis
demonstrating that rotation is a promising strategy for reducing
penetration resistance in granular media, while also highlighting a key
trade-off between force reduction and energy consumption. By examining
both intrinsic (e.g., grain and penetrator properties) and external
factors (e.g., confining pressure, depth), the findings offer meaningful
insights for designing bio-inspired self-burrowing robots and effective
penetration technologies, particularly by selecting moderate velocity
ratios, carefully tailored tip geometries, and suitable interface
friction. Future work should further clarify scaling effects,
distinguish tip- versus shaft-related resistance components, and develop
more in-depth mechanistic frameworks to explain rotation-induced
reductions of penetration resistance.

\section*{Acknowledgement}\label{acknowledgement}
\addcontentsline{toc}{section}{Acknowledgement}

This material is based upon work supported by the National Science
Foundation (NSF) under NSF CMMI 1849674, CMMI 1841574 and EEC 1449501.
Any opinions, findings, and conclusions or recommendations expressed in
this material are those of the authors and do not necessarily reflect
those of the NSF.

\newpage

\section*{List of Notations}\label{list-of-notations}
\addcontentsline{toc}{section}{List of Notations}

\noindent

\begin{table}[h]
	\begin{tabular}{lp{5.5cm}}
		$c$                                            & Constant for hyperbolic fitting, representing the ultimate reduction (-)                     \\
		$e_{\text{max}}$, $e_{\text{min}}$             & Maximum, minimum void ratios (-)                                                             \\
		$f_z^{nr}, f_z^r$                              & Shaft friction (non-rotating and rotating) per unit area at depth $z$                        \\
		$k$                                            & Constant for hyperbolic fitting, representing the reduction rate (-)                         \\
		$q_c^{nr}, q_c^r$                              & Penetration resistance (stress) on a non-rotating and rotating cone                          \\
		$u$                                            & Tangential-to-axial velocity ratio (-)                                                       \\
		$v_r$, $v_v$, $v_t$                            & Rotational, vertical, and resultant velocities  of the penetrator (m/s)                      \\
		$\mathrm{d}A$                                  & Infinitesimal area element on the cone surface (m\textsuperscript{2})                        \\
		$\mathrm{d}F_{\mathrm{fric}}$                  & Infinitesimal friction force on $\mathrm{d}A$ (N)                                            \\
		$A$                                            & Cross-sectional area at the cone base                                                        \\
		$C_c$, $C_u$                                   & Coefficients of curvature and uniformity (-)                                                 \\
		$D$                                            & Diameter of the penetrator (m)                                                               \\
		$D_{50}$                                       & Median particle size (mm)                                                                    \\
		$E$                                            & Total energy for penetration (J)                                                             \\
		$E_{nr}$, $E_r$                                & Total energy for non-rotational and rotational penetration (J)                               \\
		$E_Q$                                          & Energy consumed by penetration force (J)                                                     \\
		$E_Q^{nr}$, $E_Q^r$                            & Energy consumed by penetration force for non-rotational and rotational penetration (J)       \\
		$E_T$                                          & Energy consumed by penetration torque (J)                                                    \\
		$F_{\mathrm{fric}, z}$, $F_{\mathrm{norm}, z}$ & z-direction component of the friction force and normal force on the cone (N)                 \\
		$L$                                            & Penetration depth (m)                                                                        \\
		$L_{crit}$                                     & Critical depth marking the transition from shallow penetration to deep penetration           \\
		$Q$                                            & Total penetration force on the penetrator (N)                                                \\
		$Q_{nr}$, $Q_r$                                & Penetration force on the penetrator for non-rotaitonal and rotational penetration(N)         \\
		$Q_z$                                          & Total penetration force at depth $z$ (N)                                                     \\
		$R$                                            & Cone base radius (m)                                                                         \\
		$T$                                            & Total resistive torque on the penetrator (N·m)                                               \\
		$T_{nr}$, $T_r$                                & Total resistive torque on the penetrator for non-rotational and rotational penetration (N·m) \\
		$T_z$                                          & Total resistive torque at depth $z$ (N·m)                                                    \\
		$\alpha$                                       & Half-cone angle (rad, deg)                                                                   \\
		$\mu$                                          & Friction coefficient (-)                                                                     \\
		$\omega$                                       & Angular velocity (rad/s)                                                                     \\
		$\Phi_c(u)$, $\Phi_s(u)$                       & Rotational influence factor for the cone tip and shaft (-)                                   \\
		$\sigma_c^r$                                   & Normal stress on the rotating cone (kPa)                                                     \\
		$\sigma_{sz}^{nr}, \sigma_{sz}^{r}$            & Normal stresses on the shaft surface (non-rotating and rotating) at depth $z$                \\
		$\tau_r$                                       & Shear (frictional) stress on the cone (kPa)                                                  \\
		$\Theta$                                       & Total rotational angle (rad)                                                                 \\
		$r, \theta, z$                                 & Radial, circumferential, and vertical coordinates in cylindrical system                      \\
		$\hat{r}, \hat{\theta}, \hat{z}$               & Corresponding unit vectors in the $r$, $\theta$, and $z$ directions                          \\
		$\hat{s}, \hat{n}$                             & Unit vector along and outward orthogonal to the cone’s slanted generatrix                    \\
		$\mathbf{v}(r)$, $\mathbf{v}_{\mathrm{tang}}$  & Velocity of a point on the cone at radius $r$ and its projection on the tangent plane        \\
	\end{tabular}
\end{table}

\section*{References}\label{references}
\addcontentsline{toc}{section}{References}

\renewcommand{\bibsection}{}
\bibliography{geotechnique-cited.bib}

\phantomsection\label{appendix-count}
\section*{Appendix}\label{appendix}
\addcontentsline{toc}{section}{Appendix}

\begin{figure}

	\centering{

		\pandocbounded{\includegraphics[keepaspectratio]{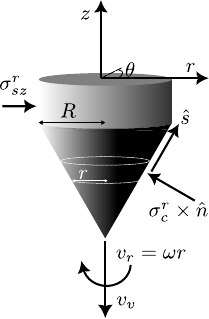}}

	}

	\caption{\label{fig-theoretical-schematic}Schematic for the theoretical
		analysis}

\end{figure}%

Assuming the soil properties are homogeneous at the scale of the cone,
closed-form expressions for the penetration forces on the cone and
shaft, both with and without rotation, can be derived.

\subsection*{Rotary Cone Penetration
	Resistance}\label{rotary-cone-penetration-resistance}
\addcontentsline{toc}{subsection}{Rotary Cone Penetration Resistance}

As illustrated in Figure~\ref{fig-theoretical-schematic}, we consider a
cone penetrator in a cylindrical coordinate system \((r, \theta, z)\),
where \(r\) is the radial coordinate, \(\theta\) is the circumferential
coordinate, and \(z\) is the vertical coordinate. The corresponding unit
vectors at any point on the cone surface are \(\hat{r}\),
\(\hat{\theta}\), and \(\hat{z}\).

We define the unit vector \(\hat{s}\) along the slanted generatrix of
the cone and its outward orthogonal unit vector \(\hat{n}\) as:

\[
	\hat{s} = \cos\alpha\hat{z} + \sin\alpha\hat{r}, \quad
	\hat{n} = \sin\alpha\hat{z} - \cos\alpha\hat{r}
\]

where \(\alpha\) is the half-cone angle.

At a radius \(r\), the cone's velocity is given by

\[
	\mathbf{v}(r) = -v_v\hat{z} + \omega r\hat{\theta}.
\]

where \(v_v\) is the vertical penetration speed and \(\omega\) is the
angular velocity of rotation.

Because friction acts along the tangent plane, we project
\(\mathbf{v}(r)\) onto that plane:

\[
	\mathbf{v}_{\mathrm{tang}}(r)
	=
	\mathbf{v}(r)
	-
	\left[\mathbf{v}(r)\cdot\hat{n}\right]\hat{n}
\]

Assume that the normal stress \(\sigma_c^r\) on the rotating cone is
uniform, and let \(\tau_r = \mu \sigma_c^r\) be the shear (frictional)
stress, where \(\mu\) is the friction coefficient. Over an infinitesimal
ring area \(\mathrm{d}A\) on the cone surface, the frictional force is

\[
	\mathrm{d}F_{\mathrm{fric}}
	=
	\tau_r \mathrm{d}A
	=
	\mu \sigma_c^r \mathrm{d}A
	=
	\mu \sigma_c^r
	\frac{2\pi r \mathrm{d}r}{\sin\alpha}.
\]

The corresponding differential friction force in the \(z\)-direction is
obtained by taking the component of \(\mathrm{d}F_{\mathrm{fric}}\) in
the direction of \(\hat{z}\) based on the velocity vector on the tangent
plane:

\[
	\mathrm{d}F_{\mathrm{fric},z}
	=
	\mathrm{d}F_{\mathrm{fric}}
	\times
	\frac{\mathbf{v}_{\mathrm{tang}}}{\|\mathbf{v}_{\mathrm{tang}}\|}
	\cdot \hat{z}
\]

Putting this together leads to:

\[
	\mathrm{d}F_{\mathrm{fric},z} =
	\frac{2\pi \mu \sigma_c^r v_v \cos^2 \alpha}{\sin\alpha}
	\frac{r \mathrm{d}r}{
		\sqrt{(\omega r)^2 + v_v^2 \cos^2\alpha}
	}.
\]

By integrating from \(r=0\) to \(r=R\), the total friction force in the
\(z\)-direction is

\begin{align}
	F_{\mathrm{fric},z}
	 & = \int_{0}^{R} dF_{\mathrm{fric},z}                           \\
	 & = \frac{2\pi R^2 \mu \sigma_c^r \cos^2\alpha}{\sin\alpha u^2}
	\left(\sqrt{u^2 + \cos^2\alpha} - \cos\alpha\right),
\end{align} where \(u = \frac{\omega R}{v}\) is a dimensionless velocity
ratio.

Similarly, the normal force in the \(z\)-direction can be derived to be

\begin{align}
	F_{\mathrm{norm},z}
	 & = \int_{0}^{R} \mathrm{d}F_{\mathrm{norm},z}                  \\
	 & = \int_{0}^{R} \sigma_c^r \times \hat{n} \cdot \hat{z} \times
	\frac{2\pi\,r\,\mathrm{d}r}{\sin\alpha}                          \\
	 & = \pi R^2 \sigma_c^r
\end{align}

Hence, the total penetration resistance on the rotating cone is

\begin{equation}\phantomsection\label{eq-qcr}{
		\begin{aligned}
			q_c^r
			 & = \frac{F_{\mathrm{norm},z} + F_{\mathrm{fric},z}}{\pi R^2} \\
			 & = \sigma_c^r \left[1 + \mu \cot\alpha \Phi_c(u)\right]
		\end{aligned}
	}\end{equation}

where

\begin{equation}\phantomsection\label{eq-phi-cu}{
		\Phi_c(u) =\frac{2\cos\alpha}{u^2} \left(\sqrt{u^2 + \cos^2\alpha} - \cos\alpha\right)
	}\end{equation}

is a rotational influence factor for the cone varying from \(0\) (when
\(u\to 0\)) to \(1\) (when \(u\to\infty\)).

When \(u\to 0\) (i.e., \(\omega=0\)), there is no rotation, and the
cone's penetration resistance is

\begin{equation}\phantomsection\label{eq-qcnr}{
	q_c^{nr}
	= \sigma_c^{nr} (1 + \mu \cot\alpha).
	}\end{equation}

\subsection*{\texorpdfstring{Total Penetration Force Ratio,
\(Q_r/Q_{nr}\)}{Total Penetration Force Ratio, Q\_r/Q\_\{nr\}}}\label{total-penetration-force-ratio-q_rq_nr}
\addcontentsline{toc}{subsection}{Total Penetration Force Ratio,
\(Q_r/Q_{nr}\)}

For the shaft portion of the penetrator at a depth \(z\), the
non-rotating (\(nr\)) and rotating (\(r\)) resistances per unit area
are, respectively,

\begin{equation}\phantomsection\label{eq-fz}{
	f_z^{nr} = \sigma_{sz}^{nr}\mu, \quad
	f_z^r = \sigma_{sz}^r \mu \Phi_s(u).
	}\end{equation}

Here, \(\sigma_{sz}^{nr}\) and \(\sigma_{sz}^{r}\) are the normal
stresses on the shaft surface at depth \(z\) for non-rotating and
rotating cases, respectively, and

\begin{equation}\phantomsection\label{eq-phi-su}{
		\begin{aligned}
			\Phi_s(u)
			 & = \frac{v_v}{\sqrt{v_v^2 + (\omega R)^2}} \\
			 & = \frac{1}{\sqrt{1 + u^2}}
		\end{aligned}
	}\end{equation}

is a rotational influence factor for the shaft.

The total non-rotating and rotating penetration forces are:

\begin{equation}\phantomsection\label{eq-Qnr}{
		\begin{aligned}
			Q_{nr}
			 & = q_c^{nr}\,A + 2\pi R \int_0^z f_z^{nr} z\,dz \\
			 & = \sigma_c^{nr}(1 + \mu \cot\alpha) A
			+ \mu \int_0^z \sigma_{sz}^{nr} z\,dz
		\end{aligned}
	}\end{equation}

\begin{equation}\phantomsection\label{eq-Qr}{
		\begin{aligned}
			Q_r
			 & = q_c^r\,A + 2\pi R \int_0^z f_z^r z\,dz        \\
			 & = \sigma_c^r [1 + \mu \cot\alpha \,\Phi_c(u)] A
			+ \mu \Phi_s(u) \int_0^z \sigma_{sz}^{nr} z\,dz.
		\end{aligned}
	}\end{equation}

where \(A\) is the cross-sectional area at the cone base. Consequently,
the ratio of these forces is

\begin{equation}\phantomsection\label{eq-analytical-ratio}{
	\frac{Q_r}{Q_{nr}}
	= \frac{
	\sigma_c^r [1 + \mu \cot\alpha \,\Phi_c(u)] A
	+ \mu \Phi_s(u) \int_0^z \sigma_{sz}^{nr} z\,dz
	}{
	\sigma_c^{nr}(1 + \mu \cot\alpha) A
	+ \mu \int_0^z \sigma_{sz}^{r} z\,dz
	}.
	}\end{equation}

In the limit as \(u \to \infty\), the ratio becomes

\begin{equation}\phantomsection\label{eq-ratio-limit}{
		\lim_{u \to \infty} \frac{Q_r}{Q_{nr}}
		= \frac{
			\sigma_c^r A
		}{
			\sigma_c^{nr}(1 + \mu \cot\alpha) A
			+ \mu \int_0^z \sigma_{sz}^{nr} z\,dz
		}.
	}\end{equation}

\bibliographystyle{Geotech} 

\end{document}